\documentclass{article}

    \PassOptionsToPackage{numbers, compress}{natbib}
    \usepackage[final]{neurips_data_2021}

\usepackage[utf8]{inputenc} % allow utf-8 input
\usepackage[T1]{fontenc}    % use 8-bit T1 fonts
\usepackage{hyperref}       % hyperlinks
\usepackage{url}            % simple URL typesetting
\usepackage{booktabs}       % professional-quality tables
\usepackage{amsfonts}       % blackboard math symbols
\usepackage{nicefrac}       % compact symbols for 1/2, etc.
\usepackage{microtype}      % microtypography
\usepackage{xcolor}         % colors
\usepackage{caption}
\usepackage{subcaption}
\usepackage{graphicx}

\title{A ground-truth dataset of real security patches}

\author{%
  Sofia Reis \\
INESC-ID \& IST/U.Lisbon, Portugal \\
  \texttt{sofia.o.reis@tecnico.ulisboa.pt} \\
   \And
   Rui Abreu \\
   INESC-ID \& FEUP, Portugal \\
   \texttt{rui@computer.org} \\
}

\begin{document}

\maketitle

\begin{abstract}
Training machine learning approaches for vulnerability 
    identification and producing reliable tools to assist 
    developers in implementing quality software---free of 
    vulnerabilities---is challenging due to the lack of 
    large datasets and real data. Researchers have been looking at 
    these issues and building datasets. However, 
    these datasets usually miss natural language artifacts and programming language
    diversity. We scraped the entire CVE details database
    for GitHub references and augmented the data with 3 security-related datasets.
    We used the data to create a ground-truth dataset of 
    natural language artifacts (such as commit messages, 
    commits comments, and summaries), meta-data and code changes. Our dataset
    integrates a total of 8057 security-relevant commits---the equivalent to 
    5942 security patches---from 1339 different projects spanning 146 
    different types of vulnerabilities and 20 languages. 
    A dataset of 
    110k non-security-related commits is also provided. 
    Data and scripts are all available on GitHub.
    Data is stored in a .CSV file. Codebases can be downloaded using our 
    scripts. Our dataset is a valuable asset to answer research questions 
    on different topics
    such as the identification of security-relevant information using NLP models; 
    software engineering and security best practices; and, vulnerability detection 
    and patching; and, security program analysis.
\end{abstract}

\section{Introduction}\label{sec:intro}

Research surrounding security patches understanding~\cite{10.1145/3133956.3134072}, 
identification~\cite{Li_2018,ponta2018metadata,6956589,zhou2019devign,
10.1145/1394504.1394505,10.1145/1052883.1052895,976940,1624016,10.1145/3106237.3117771} and 
automated repair~\cite{8835226,10.1145/1229285.1267001} are hot topics in 
software security. However, empirically validating and reproducing these approaches is 
challenging due to the lack of widely accepted and
easy-to-use databases of real vulnerabilities~\cite{10.1145/2610384.2628055}.

Accessing this kind of data is not an easy task for several reasons:
$1$) many security patches are not publicly available; $2$) developers
do not use MITRE's convention (e.g., CVE-$2017$-$4971$) to identify  
vulnerabilities in commit messages; $3$) 
commit messages are poorly written and do not provide enough/any information.
Previous research has shown that it is possible to find 
more vulnerability-fixing commits either using keywords on commits 
messages~\cite{reis2017secbench,10.4018/IJSSE.2017070101}
or training and applying machine learning algorithms~\cite{10.1145/3106237.3117771,
10.1145/1052883.1052895,sawadogo2020learning}.

With our data, we aim to provide a ground-truth dataset---real security patches
previously validated by experts---capable of supporting research using natural language, code changes, and codebases.
We scraped the Common Vulnerabilities Exposures (CVE) Details database~\cite{cvedetails}
for references to GitHub commits, i.e., code changes used to patch CVEs. 
For each CVE, we collected CVE ID, CWE ID, Vulnerability Type, CVE Severity Score,
Summaries and GitHub references. We augmented this data with other $3$ datasets that also 
contain vulnerabilities and the URL links to security patches: Secbench~\cite{reis2017secbench,10.4018/IJSSE.2017070101},
Pontas et al.~\cite{8816802} and Big-Vul~\cite{10.1145/3379597.3387501}. We crossed the data 
from datasets with the descriptive data obtained from CVE Details to complete the information 
regarding each CVE. After merging
and cleaning the data (e.g., remove duplicates and links for branches already
commits ahead of the patch version), we collected the metadata
of each GitHub commit: message commit, author, date, comments, files, code changes and 
commit parents. 

Our dataset
integrates a total of $8057$ security-relevant commits---the equivalent to 
$5942$ security patches---from $1339$ different projects spanning $146$ different types of vulnerabilities (CWEs) 
and $20$ languages. We provide natural language artifacts 
and code changes. Codebases can be pulled/downloaded using our scripts.
For machine learning purposes, we also provide a dataset
of $110161$ security-unrelated commits. The negative dataset can be extended or even built
from scratch with our scripts. Our dataset is a valuable asset to answer research questions on different topics
such as detection of security-relevant information; commits classification; 
software engineering and security best practices; 
understand security patches, their impact, and risks; vulnerability detection 
and patching; and generic security program analysis.

\textit{\textbf{Why is our dataset important?}} Our dataset is different
from previous work in the following points:

\textbf{Secbench (2017)}~\cite{reis2017secbench, 10.4018/IJSSE.2017070101}
is a dataset of $676$ single-commit security patches---the patch and vulnerable 
version are available for each case. Patches span $51$
different classes (CWEs) and $18$ languages. Data was collected by crawling $113$ different
open-source projects and applying regular expressions on commit messages to 
detect vulnerabilities fixes (e.g., \texttt{fix xss}). Our dataset integrates
more security patches, natural language artifacts, code changes 
and a negative dataset for machine learning purposes.  

\textbf{VulniOSS (2018)}~\cite{8595169} provides
results of code metrics and release data of 
$17738$ software vulnerabilities. We do not use this dataset 
because they do not provide commit id/sha(s). Our dataset instead
provides commit data, code changes, and natural language artifacts. 

\textbf{Pontas et al. (2019)}~\cite{8816802} is a manually curated
dataset of $624$ Java vulnerabilities. The authors tracked the CVE updates 
and mapped the CVE IDs manually to 
the code versions on GitHub. The dataset only provides
the commit IDs, projects, and CVE IDs. We provide a dataset
with more programming languages and more features regarding the CVEs, commits, and code changes.

\textbf{Big-Vul (2020)}~\cite{10.1145/3379597.3387501}
is a dataset of code changes and CVE summaries for $3755$ vulnerabilities 
in C/C++. Authors scraped the CVE Details database and searched  
the resulting CVE entries on open-source repositories to map the 
respective codebases. We scraped the CVE Details database as well
and found more $2224$ security-relevant commits. Our dataset
also provides cases for other programming languages and natural 
language artifacts.

The contributions of our work are the following ones:

\textbf{Ground-truth Dataset:} We organized, cleaned and 
extended previous research work by producing a larger dataset of  
vulnerabilities---validated by experts---with new features: files extension and programming languages.
The dataset contains the details of each CVE, code changes information and natural language artifacts (CVEs summaries, commits messages and commits comments). 

\textbf{Toolset:} Both the dataset and tools implemented
to scrape, clean, and manipulate the data are available in the GitHub 
repository~\cite{githubrepo}.

\section{Data Collection}

Our dataset was built using the following methodology:

\textbf{(1) CVE Details Scraping:} We scraped the entire 
CVE Details~\cite{cvedetails} website by year. For each 
year, we obtained the entire list of published CVEs
and scraped the webpage of each CVE in the list. 
We saved all the data provided on the webpage. 
However, we only integrate the following features; CVE ID, CWE ID, CVE Severity,
Summary and GitHub references. We filtered
the results to only entries with links/references to GitHub.
References not pointing
for specific versions of the repository were removed (e.g., 
\texttt{https://github.com/branch\_x\_x}) because
we were not able to ensure that the branch was still on the 
patch version. The scripts are available
in the \texttt{scraper/} folder.

\textbf{(2) Fusing Existing Datasets:} We fused $3$
datasets of security patches 
with the CVEs obtained in the previous step (\texttt{CVE Details Scraping}): Secbench~\cite{reis2017secbench,10.4018/IJSSE.2017070101},
Pontas et al.~\cite{8816802} and Big-Vul~\cite{10.1145/3379597.3387501}.
Table~\ref{tab:commits} shows the initial
distribution of commits and patches of each dataset and their license. After
fusing the data, duplicates and entries for other source code hosting websites 
were removed. We crossed the fused data with the CVE details data
to complete the CVE entries with CWEs and summaries.
The scripts are available in the \texttt{scripts/} folder.

\begin{table}[h]
    \caption{Distribution of commits before and after phase (2)} 
    \centering 
    \begin{tabular}{c c c c c c} 
    \hline
    Dataset & \multicolumn{2}{c}{Before (2)} & \multicolumn{2}{c}{After (2)} & License\\
     & Commits & Patches & Commits & Patches &\\
     \hline\hline 
    CVE details~\cite{cvedetails} & 4529 & 4183 & 2224 & 1816 & -\\
    Secbench~\cite{reis2017secbench} & 676 & 676 & 659 & 659 & MIT\\
    Pontas et al.~\cite{8816802} & 1282 & 624 & 1127 & 565 & Apache 2.0\\
    Big-Vul~\cite{10.1145/3379597.3387501} & 4529 & 3755 & 4047 & 3433 & MIT\\
\hline  
    \end{tabular}
    \label{tab:commits}
\end{table}

\textbf{(3) Extracting Meta-Data:} We used 
the GitHub API to retrieve the meta-data of each commit:
message, author, date, changed files, comments, and parents.
Each comment has an author, date, and body text. Each file
has a path, number of additions, number of deletions, 
number of changes, and status. We built a script to add $2$  
features to the dataset: extension of the files and programming languages.
We inferred programming languages from the extension files. 

\textbf{(4) Generate Negative Dataset:} We used 
the GitHub API to retrieve $50$ random security-unrelated commits per project of the positive dataset. We cleaned the commits using security-related keywords such as ``secur'', ``patch'', ``cve'', 
``vuln'', ``attack''. This dataset integrates only two columns: commit link and commit message. The scripts are available
in the \texttt{negative/} folder.

\begin{table}[h]
    \caption{Description of the positive data collected (security patches)} 
    \centering 
    \begin{tabular}{| p{3.5cm} | p{9cm} |} 
    \hline
    \multicolumn{2}{|c|}{\textbf{CVE Details}}\\
    \hline
    Column Name (.CSV) & Description\\
    \hline
    cve\_id & The common vulnerabilities and exposures identifier. \\\hline
    project & GitHub project name.\\\hline
    sha & Commit key or identifier of the version in the project repository.\\\hline
    cwe\_id & The common weakness enumeration identifier.\\\hline
    score & Severity score of the vulnerability.\\
    \hline\hline
    \multicolumn{2}{|c|}{\textbf{Code Changes}}\\
    \hline
    Column Name (.CSV) & Description\\
    \hline
    files & Set of files changed by the patch. Information schema: path, additions, deletions, changes, status.\\\hline    
    language & Programming Language.\\\hline  
    github & Commit link.\\\hline 
    parents & Commit keys for previous software version.\\\hline
    date & Date of the changes.\\\hline
    author & Author of the changes.\\\hline
    ext\_files & Extension of the files.\\\hline
    lang & Programming Language.\\\hline
    \hline
    \multicolumn{2}{|c|}{\textbf{Natural Language Artifact}}\\
    \hline
    Column Name (.CSV) & Description\\
    \hline
    summary & Summary of the vulnerability.\\\hline    
    message & Commit message.\\\hline
    comments & Comments set. Information schema: author, date and body.\\
\hline  
    \end{tabular}
    \label{tab:data}
\end{table}

\section{Dataset Description}

In total, we collected $8057$ security-relevant commits---the equivalent to 
$5942$ security patches---from $1339$ different projects spanning $146$ different types 
of vulnerabilities (CWEs) and $20$ languages. The dataset integrates $16$ different 
columns of data for each commit. Table~\ref{tab:data} describes
in more detail each feature of the dataset. We provide scripts 
to retrieve the CVE Details data (\texttt{scraper} folder); to fuse, clean and manipulate the data (\texttt{scripts} folder); to augment data by adding features such as the extension files and 
the programming language in which the commits were implemented the data (\texttt{scripts/add\_features.py} script); and, scripts to generate the negative dataset, i.e., the dataset with non-security related commits (\texttt{negative} folder). Jupyter notebooks are also provided to show the statistics presented in this section (\texttt{notebooks} folder).

\begin{figure}
    \centering
    \includegraphics[width=0.6\linewidth]{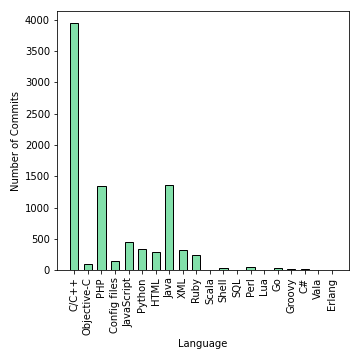}
    \caption{Distribution of commits per programming language. The top-$3$ of most prevalent
programming languages are \texttt{C/C++} (3944 commits), \texttt{Java} (1369 commits) and \texttt{PHP} (1350 commits).}
    \label{fig:lang}
\end{figure}

Figure~\ref{fig:lang} shows the number of commits implemented in each language
for the $20$ languages available in our dataset. The top-$3$ of most prevalent
programming languages are \texttt{C/C++} (3944 commits), \texttt{Java} (1369 commits) and \texttt{PHP} (1350 commits).
Security-relevant commits are less prevalent amongst \texttt{Erlang} (3), \texttt{Lua} (5) and \texttt{Vala} 
(3)---less than $15$ commits were obtained per language.

\begin{figure}
    \centering
    \includegraphics[width=0.6\linewidth]{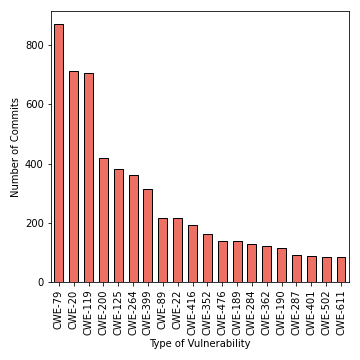}
    \caption{Distribution of commits for the top-20 of vulnerabilities (CWEs). The dataset comprises commits to patch 
$146$ different types of CWEs.}
    \label{fig:cwe}
\end{figure}

Figure~\ref{fig:cwe} shows the amount of commits distributed by the top-$20$ 
of vulnerabilities (CWEs). Our dataset comprises commits to patch 
$146$ different types of CWEs. The most common commits patch vulnerabilities 
such as the \textit{CWE-79: Improper Neutralization 
of Input during Web} ($870$ commits), \textit{CWE-20: Improper Input Validation} ($712$ commits),
\textit{CWE-119: Improper Restriction of Operations within the Bound of a Memory Buffer} ($705$ commits), 
\textit{CWE-200: Exposure of Sensitive Information to an Unauthorized Actor} ($419$ commits), 
\textit{CWE-125: Out-of-bounds Read} ($380$ commits). These represent $38.1\%$ of the entire
dataset.

\begin{figure}
    \centering
    \includegraphics[width=0.5\linewidth]{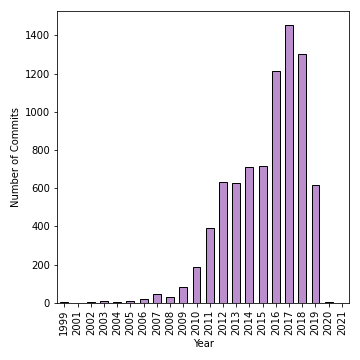}
    \caption{Distribution of commits per year (from 1999 to 2021). The number of commits between $2016$ and $2019$ are the most predominant ones. }
    \label{fig:year}
\end{figure}

Figure~\ref{fig:year} shows the number of commits used to patch
software vulnerabilities per year. The security-relevant commits of this dataset 
were published from $1999$ until $2021$. The number of security-relevant commits between $2016$ and $2019$ are the most predominant ones. 

\begin{table}[h]
    \caption{Files and code changes descriptive data} 
    \centering 
    \begin{tabular}{| p{3cm} | p{2cm} | p{2cm} |} 
        \hline

        \multicolumn{3}{|c|}{\textbf{Files}}\\\hline
        Code files & 26032 & 76.87\%\\\hline
        Test files & 6770  & 19.99\%\\\hline
        Changelogs/News &  933 & 2.75\% \\\hline
        Readme & 131 & 0.39\%\\\hline
        Total of Files & 33866 & 100\%\\\hline
        \hline
        \multicolumn{3}{|c|}{\textbf{Code Churn}}\\\hline
        Additions ({\color{green}+++})& 1113450  & 74.60\% \\\hline
        Deletions ({\color{red}- - -})& 379203  & 25.40\%\\\hline
        Total of Changes & 1492653  & 100\%\\\hline
    \end{tabular}
    \label{tab:files}
\end{table}

Table~\ref{tab:files} shows the distribution of files and code changes
of the positive dataset. A total of $8057$ security-related commits were performed
across $33866$ different files: $76.87\%$ are code files, $19.99\%$ are test
files, $2.75\%$ are changelogs and news files; and, $0.39\%$ are readme files.
To patch the total of $5942$ security patches integrated into this dataset,
$1113450$ lines of code were added, and $379203$ lines of code were
deleted. In total, more than $1$M of changes were performed.

\begin{figure*}
    \begin{subfigure}{.33\textwidth}
      \centering
      \includegraphics[width=.9\linewidth]{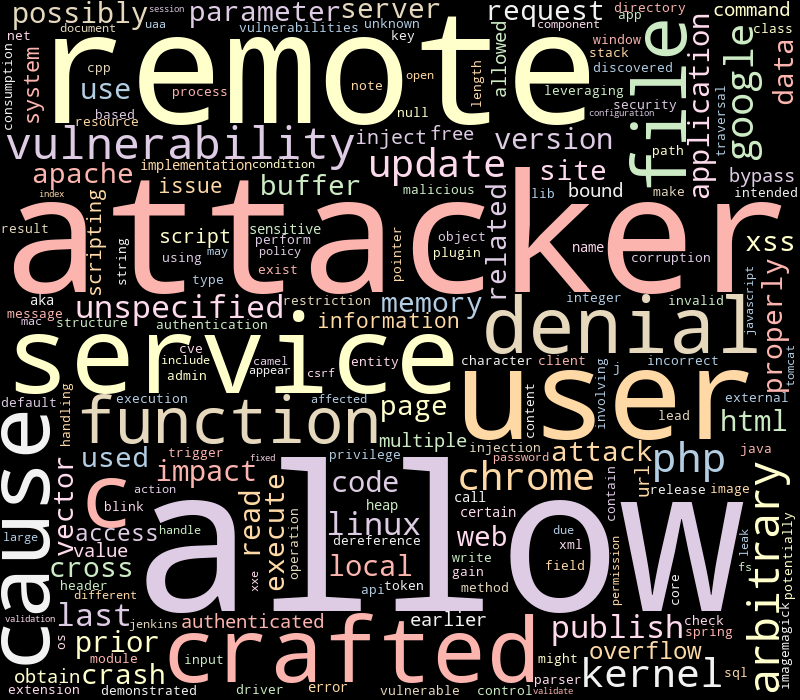}
      \caption{CVE Summary}
      \label{fig:cve}
    \end{subfigure}
    \begin{subfigure}{.33\textwidth}
      \centering
      \includegraphics[width=.9\linewidth]{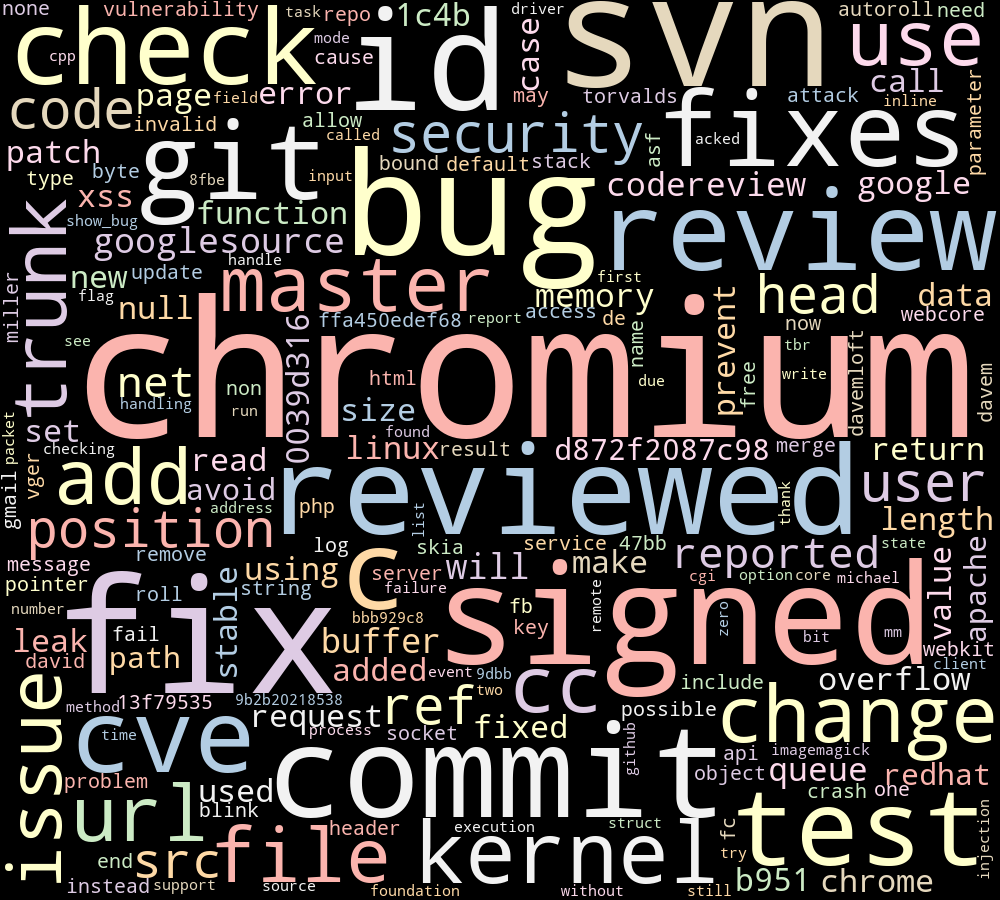}
      \caption{Commit Message}
      \label{fig:msg}
    \end{subfigure}
    \begin{subfigure}{.33\textwidth}
        \centering
        \includegraphics[width=.9\linewidth]{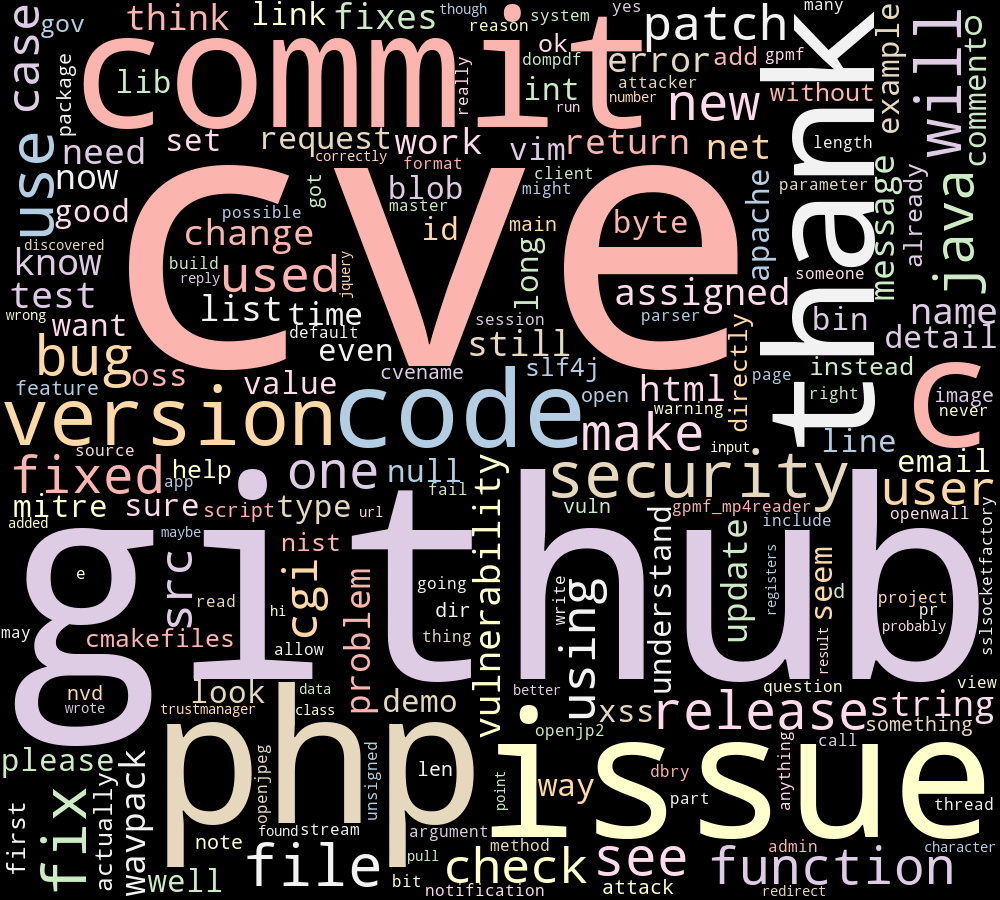}
        \caption{Developers Comments}
        \label{fig:com}
      \end{subfigure}
    \caption{Wordclouds for each type of natural language artifact: (a) CVE Summary (col: \texttt{summary}), (b) Commit Message (col: \texttt{message}), and, (3) Developers Comments (col: \texttt{comments})}
    \label{fig:wordclouds}
\end{figure*}

Figure~\ref{fig:wordclouds} shows wordclouds for the different 
natural language artifacts: CVE summaries (Figure ~\ref{fig:cve}), 
messages (Figure ~\ref{fig:msg}) and comments (Figure ~\ref{fig:com}). Each wordcloud
shows the importance of the words used in our artifacts. 
CVE summaries are brief explanations of vulnerabilities and possible
attacks---words such as ``attacker'', ``remote'',
``service'', ``allows'' and ``user'' are the most common. 
While security-relevant commit messages usually describe an action 
such as fixing a vulnerability. Figure ~\ref{fig:msg}
supports this by showing words such as ``fix'', ``issue'', ``review'',
and ``change''. Comments are usually performed by developers, project contributors
or users asking for information on the vulnerability/code. The most important words are ``issue'', ``commit'', ``CVE''
``thank'' and ``github''. Overall, wordclouds reflect the natural language artifacts
goals.

A dataset of security-unrelated commits (negative dataset) was generated from the projects integrated in the dataset of security patches. For each project in the positive dataset, we collected $50$ random commits. Non-security related
keywords were leveraged to clean the dataset. In total, we selected $110161$ random non-security related commits from the projects in the positive dataset. Each entry of the negative dataset represents a non-security-related commit. For each commint, the dataset integrates 1) the github link for the code changes and 2) the commit message of the change. 

Both datasets are available on our GitHub 
repository in the comma-separated values (CSV) format (\texttt{dataset} folder). Scripts to 
download the codebases of the commits are provided. The jupyter 
notebooks used to report this analysis are also available.  

\section{Future research and Improvements}

One of the issues in the field of security program analysis 
is the lack of easy-to-use, well-organized and diverse 
datasets of security patches. Our dataset provides 
a larger dataset considering commits for $20$ programming languages, 
$146$ types of vulnerabilities and $3$ different types of artifacts: meta-data,
natural language data and codebases. This dataset can be used in several 
vulnerability related research topics such as:

\textbf{Identification and classification of security-relevant commits.} 
This is a recent hot topic in the software engineering field. Machine learning models to identify 
and classify automatically this kind of commits are being created~\cite{10.1145/3106237.3117771, sawadogo2020learning}.
These models aim to fasten software releases and notify developers when new vulnerabilities 
are reported. Our natural language artifacts can be used to train deep learning approaches based
on text mining and natural language processing to detect this type of commits.

\textbf{Vulnerability detection and patching.} Over the past $2$ decades, a lot of research 
as been developed in the vulnerability detection field~\cite{Li_2018,ponta2018metadata,6956589,zhou2019devign,
10.1145/1394504.1394505,10.1145/1052883.1052895,976940,1624016,10.1145/3106237.3117771}. More recently, 
research in vulnerability patching has emerged~\cite{8835226,10.1145/1229285.1267001}. However, 
many of these techniques are not well tested due to the lack of code samples and do not cover
some types of vulnerabilities. 
Deep learning approaches have a huge potential to tackle the later issue, but due to the lack 
of data to well train and test the models, these approaches do not succeed. We provide a large
dataset of real data capable of boosting research in these fields.

\textbf{Understand the impact and risks of security patches.} Researchers performed
a large scale empirical study to understand security patches 
characteristics~\cite{10.1145/3133956.3134072}. However, there is a low understanding on 
the impact and risks of security patches on software. Our scripts allow the user to download 
the codebases of the commits, previous commit(s), and the diff between both.
With code-centric information, researchers can better understand 
security patches and their features. New neural networks can be trained with this data
to detect and patch
vulnerabilities. This kind of data can also be used to create models using software properties to predict 
the risk of patches on companies and software development.

\textbf{Explore if the best practices of software engineering and security are being used.}
Code evolves and teams change. Assessing the quality of commits and reports may be crucial
to make development easier for everyone. Our natural language artifacts can be used to 
explore if best practices are being used. Code changes data can also be leveraged to better
understand if the best practices of development are being used. 

\section{Limitations and Challenges}\label{sec:limitations}

Our dataset integrates security patches without a CVE assigned. Thus, 
we were not able to get the CVE Details meta-data for these cases:
$10\%$ of the rows are missing descriptive data, e.g., the vulnerability identifier (CVE ID), 
summary, type of vulnerability (CWE ID), and score. On the other 
hand, our methodology 
to infer the programming language from the extension files involved
in the commits failed on classifying less than $5\%$ of the entire dataset.
Thus, some commits might not have a programming language associated.
Both issues will be improved/addressed in the future. 

\section{Conclusions}

We fused $4$ different security-related data sources
and created a ground-truth dataset of 
natural language artifacts, meta-data, and code changes. Our dataset
integrates a total of 8057 security-relevant commits---the equivalent to 
5942 security patches---from 1339 different projects 
spanning 146 different types of vulnerabilities 
and 20 languages. In addition, this dataset provides a 
negative dataset of 110k non-security-related commits. 
Data and scripts are available on GitHub.
Data is stored in a .CSV file and codebases can be download using our 
scripts. Our goal is to extend this dataset in the future by crossing the entire CVEs list with github information. An interactive website to 
navigate the data is under development. Our dataset is a valuable asset to 
answer research questions on different topics
such as the identification of security-relevant commits; 
software engineering and security best practices; 
understand security patches, their impact, and risks; and, vulnerability detection 
and patching.

\section{Acknowledgments}

This material is based upon work supported by the 
scholarship number SFRH/BD/143319/2019 from Fundação para a 
Ciência e Tecnologia (FCT). 

\bibliographystyle{acm}
\bibliography{neurips}

% %%%%%%%%%%%%%%%%%%%%%%%%%%%%%%%%%%%%%%%%%%%%%%%%%%%%%%%%%%%%

%%%%%%%%%%%%%%%%%%%%%%%%%%%%%%%%%%%%%%%%%%%%%%%%%%%%%%%%%%%%

\end{document}